\documentclass[journal]{IEEEtran}
\usepackage{amsmath,amsfonts}
\usepackage{array}
\usepackage[ruled, linesnumbered, commentsnumbered, longend]{algorithm2e}
\usepackage{textcomp}
\usepackage{stfloats}
\usepackage{url}
\usepackage{optidef}
\usepackage{verbatim}
\usepackage{graphicx}
\usepackage{subcaption}
\usepackage{relsize}
\usepackage{cite}
\usepackage{bbm}
\usepackage{graphicx, color, xcolor, colortbl}
\newcolumntype{L}[1]{>{\raggedright\let\newline\\\arraybackslash\hspace{0pt}}m{#1}}
\newcolumntype{C}[1]{>{\centering\let\newline\\\arraybackslash\hspace{0pt}}m{#1}}
\newcolumntype{R}[1]{>{\raggedleft\let\newline\\\arraybackslash\hspace{0pt}}m{#1}}
\usepackage[shortlabels, inline]{enumitem}

\newcommand{\eg}{\textit{e.g.,}\xspace}
\newcommand{\ie}{\textit{i.e.,}\xspace}
\begin{document}

\title{Federated Deep Reinforcement Learning for Open RAN Slicing in 6G Networks
}

\author{Amine Abouaomar,~\IEEEmembership{Member,~IEEE,} Afaf Taik,~\IEEEmembership{Member,~IEEE}, Abderrahime Filali,~\IEEEmembership{Member,~IEEE,} and Soumaya Cherkaoui,~\IEEEmembership{Senior~Member,~IEEE,}

    \thanks{Amine Abouaomar, Abderrahime Filali and Soumaya Cherkaoui are with the Department of Computer and Software Engineering, Polytechnique Montreal (e-mails: amine.abouaomar@polymtl.ca abderrahime.filali@polymtl.ca, soumaya.cherkaoui@polymtl.ca)}
    \thanks{Afaf Taïk is with Université de Sherbrooke (e-mails: afaf.taik@usherbrooke.ca)}
}

\markboth{}%
{Filali \MakeLowercase{\textit{et al.}}: Federated Deep Reinforcement Learning for Open RAN Slicing in 6G Networks}

\maketitle
\begin{abstract}
Radio access network (RAN) slicing is a key element in enabling current 5G networks and next-generation networks to meet the requirements of different services in various verticals. However, the heterogeneous nature of these services' requirements, along with the limited RAN resources, makes the RAN slicing very complex. Indeed, the challenge that mobile virtual network operators (MVNOs) face is to rapidly adapt their RAN slicing strategies to the frequent changes of the environment constraints and service requirements. Machine learning techniques, such as deep reinforcement learning (DRL), are increasingly considered a key enabler for automating the management and orchestration of RAN slicing operations. Nerveless, the ability to generalize DRL models to multiple RAN slicing environments may be limited, due to their strong dependence on the environment data on which they are trained. Federated learning enables MVNOs to leverage more diverse training inputs for DRL without the high cost of collecting this data from different RANs. In this paper, we propose a federated deep reinforcement learning approach for RAN slicing. In this approach, MVNOs collaborate to improve the performance of their DRL-based RAN slicing models. Each MVNO trains a DRL model and sends it for aggregation. The aggregated model is then sent back to each MVNO for immediate use and further training. The simulation results show the effectiveness of the proposed DRL approach.
\end{abstract}
\begin{IEEEkeywords}
Federated Deep Reinforcement Learning, Open RAN, RAN Slicing, Federated Learning, Reinforcement Learning, B5G, 6G.
\end{IEEEkeywords}

\section{Introduction}
Beyond 5G and 6G are expected to support the explosive growth of data traffic generated by a huge number of connected devices \cite{jiang2021road}. These devices enable advanced services to multiple vertical industries with diversified quality of service (QoS) requirements. Therefore, the traffic volume in these next-generation networks will grow exponentially carrying a huge amount of data. This will increase the demand for bandwidth to provide the connectivity required to transfer this data efficiently and rapidly. Accordingly, these next-generation networks will entail more stringent requirements than current 5G networks. To support all these services, mobile network operators (MNOs) are required to provide an adequate network infrastructure. To achieve this in a high-performance and cost-effective manner, MNOs rely on several cutting-edge technologies, such as network slicing (NS). NS allows the design of several logically independent networks, so-called network slices, which operate on a common physical infrastructure \cite{filali2020survey}. In particular, radio access network (RAN) slicing consists in partitioning the RAN resources to create various RAN slices, each tailored and dedicated to meet the requirements of a specific service. An MNO owns the physical RAN resources, including radio resources, and leases them to mobile virtual network operators (MVNOs) to deploy RAN slices based on their offered services. 

The allocation of radio resources to users is an extremely intricate operation for MVNOs. This is mainly due to the radio resources' scarcity and the heterogeneous QoS requirements of their services \cite{song2021deep}. To overcome these challenges, several efforts have been devoted to formulating such a RAN slicing problem using optimization techniques and solving it with heuristics \cite{motalleb2022resource,ravi2021ran}. However, these approaches may not align with zero-touch network perspectives since they may be unable to quickly adapt to dynamic and permanent changes in the RAN environment. Machine learning (ML) techniques, specifically deep reinforcement learning (DRL) algorithms, can address this issue by bringing more automation to the management of RAN slicing operations. The stochastic RAN environment factors, such as the density of users, user requirements, and wireless channel transmission conditions have a major impact on the accuracy of the DRL models, which decreases the performance of radio resource allocation to the users. Indeed, when an MVNO builds its DRL resource allocation model using training datasets related only to its users' behavior and its surrounding environment, the accuracy of the DRL model may be limited. To benefit from a diversified dataset, MVNOs could collaborate by sharing their data with each other to provide a diverse and high-quality dataset. However, MVNOs are often competing entities and are unlikely to be willing to share their data for privacy issues. Federated learning (FL) has emerged as a promising alternative to data sharing, as it promotes privacy-preserving collaborative training through sharing model updates instead \cite{konevcny2016federated,kairouz2021advances}.

FL is a cooperative learning approach in which multiple collaborators, MVNOs in our case, train an ML model using their private datasets, and then send their trained models to an aggregation entity to build a global model. The aggregation entity returns this model to all collaborators for utilization or further training. Thus, FL enables MVNOs to build a robust ML resource allocation model while maintaining data privacy since only trained models are shared. The shared experience will enable the RAN-slicing model to learn from varying scenarios, which makes it more adaptive to the environment changes. In fact, due to the unbalanced and non-independent, and identical distributions (non-i.i.d) of the users across MVNOs, alongside their varying numbers and requirements, FL becomes an attractive solution to build robust models.

To implement a FL-based RAN slicing mechanism with a high level of automation and flexibility, Open RAN (for which the O-RAN alliance has built the O-RAN architecture as the foundation for virtualized RAN on open hardware and cloud \cite{ORAN}) emerges as a revolutionary solution. O-RAN relies on openness and intelligence to provide flexible and cost-effective RAN services. Openness refers to the disaggregation of the RAN into several independent functions. Then, the use of open, standardized, and interoperable interfaces to ensure communication between these RAN functions. Thus, the need to use proprietary equipment will be reduced, which leads to more competition among vendors and further innovation. Also,  O-RAN promotes the use of more intelligence and automation in controlling the RAN operation through RAN intelligent controllers (RICs). The RICs leverage the capabilities of ML to manage the RAN resources and components such that it becomes a zero-touch network system.


The motivation behind this work is to build reliable DRL-based radio resource allocation models for MVNOs by leveraging their heterogeneous and diverse datasets while preserving data privacy and security. Accordingly, we propose an FL-based cooperative radio resource allocation mechanism for MVNOs. In this mechanism, each MVNO trains an DRL radio resource allocation model according to its users' requirements and sends the trained models to the RIC for aggregation. Then, the RIC sends back the global DRL model to each MVNO to update its local DRL models. The proposed RAN slicing approach is consistent with the zero-touch network framework since it aims to autonomously manage the allocation of radio resources to users by adapting the allocation policy to the constraints of the environment and the requirements of users. Indeed, the use of DRL enables the implementation of a closed-loop control process of the RAN slicing operations. Furthermore, our proposed RAN slicing mechanism is implemented in an O-RAN architecture that favors using more intelligence and automation in monitoring RAN operations.

 The main contributions of this paper are summarized as follows: we
\begin{enumerate*}[(i)]
    \item discuss DRL-based RAN slicing efforts and associated challenges in a multi-MVNOs environment, 
    \item design a federated DRL (FDRL) mechanism on an O-RAN architecture to improve the radio resource allocation operation of MVNOs, and
    \item  illustrate, through extensive simulations, that the proposed RAN slicing mechanism enables a better allocation of the needed radio resources to satisfy users' QoS requirements in terms of delay and data rate.
\end{enumerate*}

\section{ DRL AS A RAN SLICING ENABLER.}
\subsection{DRL-based RAN Slicing efforts}
\begin{figure*}[ht]
    \centering
    \includegraphics[width=.7\linewidth]{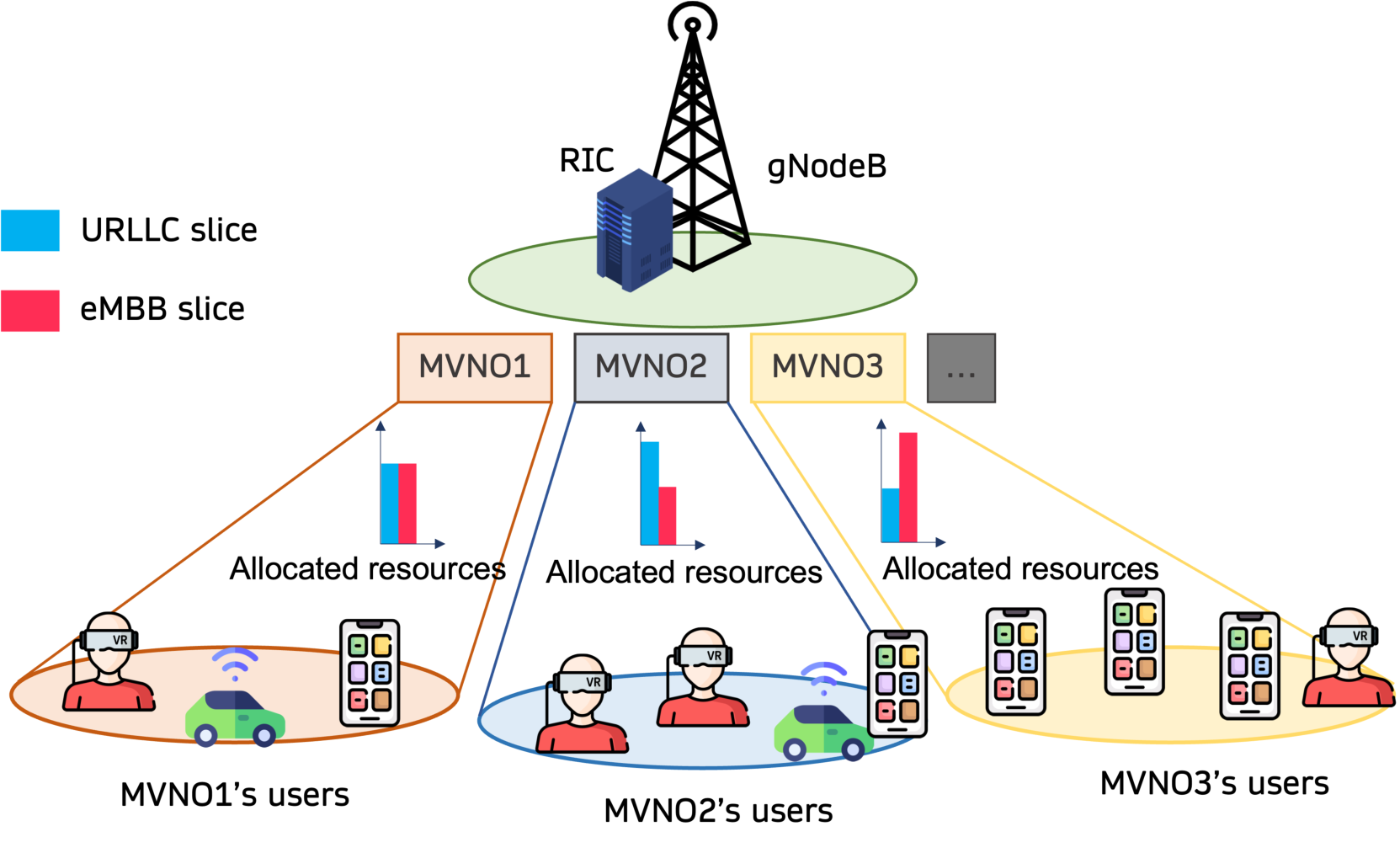}
    \caption{\label{fig:sm}An overview of the system model.}
\end{figure*}
DRL has been widely used as an efficient tool to perform RAN slicing. In \cite{setayesh2022resource}, the DRL is used to determine the configuration parameters of the RAN slices. It defines for each RAN slice the transmit power, the bandwidth, the shared bandwidth between the slices and the numerology. The work in \cite{filali2022tnse} designs a DRL-based RAN slicing mechanism used by base stations to allocate radio resources to their associated users. When the resources of a base station are not sufficient, this mechanism consists in requesting additional resources from other base stations to satisfy the QoS requirements of its users.

To improve the RAN slicing operation in a multi-MVNOs scenario, various DLR-based efforts have been presented recently. For instance, \cite{chen2022two} proposes a two-level slicing resource allocation scheme using DRL. Based on the resources received from the MNO, each MVNO allocates to its users the required radio resource to satisfy their QoS. The MVNO uses a DRL algorithm to perform resource allocation to its users in order to maximize their satisfaction rate, which is defined by the number of packets successfully received from the network. The work in \cite{wang2020utility} presents a DRL-based resource allocation algorithm to maximize the utility of MVNOs while meeting QoS requirements of RAN slices. The utility of an MVNO is defined as the gain from allocating the resource to the slice minus the cost of leasing the resource from the MNO. 

In all these approaches DRL has been used as an enabler for orchestrating RAN slicing. However, the ability to generalize DRL models to multiple RAN slicing environments may be limited, due to their strong dependence on the environment data on which they are trained. The proposed FDRL-based RAN slicing mechanism overcomes the robustness issue that can arise in approaches based only on DRL. FDRL enables MVNOs to leverage more diverse training inputs for DRL while avoiding to share the data of their RAN environments.

\subsection{RAN Slicing Challenges}
Although many approaches have shown that DRL enables efficient RAN slicing, several challenges remain open and deserve further investigation.

\subsubsection{Robustness}
The main drawback of policy-learning methods, such as DRL, is their dependence on data. Training robust DRL models requires huge amounts of data. Note that the RAN environment has a high uncertainty related to the various user requirements, dynamic wireless transmission conditions, and traffic permanent fluctuations. Accordingly, the training in such environments can suffer from the scarcity of data. Thus, there is a need for providing heterogeneous scenarios to build robust DRL-based RAN slicing models.

\subsubsection{SLA satisfaction}
In RAN slicing, SLA presents the minimum QoS requirements that should be guaranteed on a provided service. In beyond 5G/6G networks, RAN slicing is expected to deal with more fine-grained SLA. In addition, resources in the RAN environment are limited, leading to resource scarcity issues. Therefore, network operators need to design DRL algorithms that optimize resource allocation to meet SLAs and thus provide maximum service.
\vspace{-0cm}
\subsubsection{RAN slice isolation}: DRL-based RAN slicing mechanisms can ensure flexible management of RAN slices. They can rapidly instantiate or reconfigure RAN slices to meet the requirements of users. However, a fast configuration of RAN slices can be achieved at the expense of their isolation. Therefore, DRL mechanisms must consider the tradeoff between fast management and accurate isolation of RAN slices.
 
\subsubsection{Privacy and competition}
While encouraging MVNOs to collaborate to build robust slicing models is desirable, several privacy issues arise. Indeed, MVNOs have no incentives to share data related to the QoS of their users with competitors. Instead, new techniques that promote both collaboration and privacy are needed to guarantee high SLA satisfaction.

In this work, we design a FDRL mechanism to allocate radio resources in a multi-MVNOs environment. We use FL to build a robust and private DRL-based RAN slicing mechanism that satisfies MVNOs' users in terms of SLA. The robustness of the proposed mechanism is achieved since the MVNOs collaborate the train a global DRL model. The MVNOs share their models locally trained based on their scenarios to build a robust global RAN slicing model. 

\section{FDRL-Based Cooperative RAN Slicing Between MVNOs}

    
\subsection{System Model}
We consider a RIC-enabled RAN architecture with a single base station, \ie gNodeB, owned by an MNO. The BS is operating on a total bandwidth B. The MNO is responsible to serve a set of MVNOs by renting to each of them a fraction of the total bandwidth. Each MVNO has a set of users that upload their packets to the network. We consider two types of users, namely enhanced mobile broadband (eMBB) and ultra-reliable low-latency communication (URLLC).

We assume that bandwidth allocation to MVNOs has already been performed by the MNO.  An MVNO allocates to each of its users a fraction of the bandwidth pre-allocated by the MNO to satisfy its QoS requirements when uploading packets. Indeed, the goal for each MVNO is to efficiently allocate the bandwidth to ensure a high data rate for eMBB users and a low transmission delay for URLLC users. 
In this paper, we propose a FDRL-based RAN slicing mechanism for MVNOs, which collaborate to improve the performance of allocating their communication resources to their users. 


\subsection{Deep Deterministic Policy Gradient (DDPG)}
The RAN environment is highly dynamic, where traffic fluctuations are permanent and wireless transmission conditions change frequently. Thus, it is challenging for an MVNO to perform the radio resource slicing operation in a such an environment. To overcome this challenge, ML techniques such as DRL can be leveraged.
To solve the RAN slicing problem of an MVNO based on the DRL algorithm, we need to model this problem as a Markov decision process (MDP).

\textbf{MDP Formulation:}
 For each MVNO, the bandwidth allocation problem is modeled as a single-agent MDP as follows:
\begin{itemize}
    \item \textit{The State Space:} The state observed by the agent includes the type of its users and their channel gains. The type of users, \ie eMBB and URLLC, is necessary to define the requirements of each user. The channel gain between a user and the gNodeB is calculated using the state channel information collected by the gNodeB.  Since each MVNO might serve a different number of users each time, it raises a problem related to the size of the input and output of the trained model. To overcome this problem, we set the input size and output size according to the maximum number of users that an MVNO can serve at one time. In case the observed number of users is lower than this maximum user threshold, we use zero-padding. This allows us to adapt the varying number of users of each MVNO and to unify the trained model.
     \item \textit{The Action Space:} An agent has to decide which bandwidth fraction should be allocated to each of its users. The action chosen for a user is a real value between 0 and a threshold that defines the maximum bandwidth faction that a user can have.   
     \item \textit{The Reward Function:} 
     In this work, the main objective is the ensure a high data rate for the eMBB users and a low transmission delay for the URLLC users. The reward depends on the QoS provided to each of its users in terms of data rate and delay. Accordingly, the reward obtained by an agent is the total sum of the data rates achieved by the eMBB users and the inverse of the transmission delays experienced by the URLLC users. If the achieved data rate of an eMBB user is less than a minimum threshold, or if the latency experienced by an URLLC user is greater than a maximum threshold, this action will be punished by -0.1. Theses punishments will be added to the reward obtained by the agent. Moreover, if the total sum of the bandwidth fractions is greater than 1, the global action is considered as invalid, and it will be penalized by -0.05. Furthermore, in order to avoid the case where a fraction of the bandwidth is allocated to a user that does not exist, we associate this action to a punishment equal to a negative value that we add to the reward (-0.1).
\end{itemize}
DRL allows to create agents that learn from high-dimensional states, where the policy is represented as a deep neural network. DRL was first introduced through Deep-Q Networks (DQN), and was fast adopted by the research community to solve many practical decision making problems. Nonetheless, DQN is off-policy and may not perform well in environments that have high uncertainties such as wireless networks. While value-based RL algorithms like Q-learning optimize value function first then derive optimal policies, the policy-based methods directly optimize an objective function based on the rewards, which makes them suitable for large or infinite action spaces. Yet, policy-based RL might have noisy and unstable gradients. As a result, we propose to use an actor-critic based algorithm. In fact, actor-critic approaches combine strong points from both value-based and policy-based RL algorithms. Furthermore, since the fraction values are continuous, we use the deep deterministic policy gradient (DDPG) algorithm, which concurrently learns a Q-function and a policy and performs actions from a continuous space.

\textbf{DDPG algorithm:} It uses 4 neural networks: the actor network, the critic network, the actor target network, and the critic target network. For a given observed environment state, the actor chooses an action, and the critic uses a state-action Q-function to evaluate it \cite{lillicrap2015continuous}. To reduce the correlation between the training samples, DDPG uses the experience replay memory, which stabilizes its behavior. An experience is defined by the current state of the environment, the chosen action, the reward obtained, and the next state of the environment. The agent stores its experiences and then samples random mini-batches from them to train its DQN networks. The exploration policy of DDPG is performed by adding noise to the actions during the training process. The added noise enables the DDPG agent to efficiently explore its environment. We used the Ornstein–Uhlenbeck (OU) process to generate the noise values.


    
\subsection{FDRL-based RAN Slicing mechanism}
The disparity of users distributions across different geographical areas (\eg suburbs and city center), for instance, makes using the same model across all the covered areas inadequate. Moreover, the amount of data collected by each MVNO in certain areas (\eg rural areas) is fairly limited. Since it is beneficial for each MVNO to enhance its bandwidth allocation model, FL has created the opportunity for multiple MVNOs to leverage data from a broader set of users while avoiding sharing it \cite{wang2022linkslice}. In this setting, each MVNO trains the bandwidth allocation model using the interaction with its users. Then, the MVNO uploads its locally trained model for the current round to the non-RT RIC. The non-RT RIC performs the aggregation of the models using weighted sum based on each MVNO's number of users.


 The FDRL-based mechanism consists of several communication rounds. Each communication round consists of several local episodes of training, after which the MVNOs update their local models and send the updates to the non-RT RIC. 
In each step of the episode, each MVNO observes their local environment, selects an action and receives the corresponding reward.  Each transition between two states, \ie experience, is stored in the replay buffer. When a predefined number of experiences is stored, MVNOs sample random mini-batches from the replay buffer, and update the different DDPG networks. The actor network is updated through policy gradient and the critic network is updated through loss function minimization. Subsequently, the target networks are updated as well. At the end of the predetermined number of local episodes, each MVNO sends its local updated model to the non-RT RIC for aggregation. The latter collects all the local updates from the MVNOs and generate the global model using a weighted sum.

The O-RAN Alliance Working Group 2 defines in \cite{ORANWG2} three scenarios for the deployment of ML model learning and ML model inference. In this work, we opted for the scenario where the offline training process is hosted in the non-RT RIC and the online inference process is hosted in the near-RT RIC. This choice is related to the proposed FDRL-based RAN slicing mechanism. Indeed, we used a DDPG algorithm, which requires a huge amount of data to perform offline training of RAN slicing models. This offline training requires a resourceful unit with high computing and storage capabilities. Therefore, the non-RT RIC seems better suited to host the offline training of RAN slicing models. The near-RT RIC was chosen as the model inference host since 1) the radio resource allocation operation to users should be performed in a short time scale, and 2) the data needed at the execution time of the RAN slicing operation is available through the E2 interface, which connects the near-RT RIC to the RAN nodes.
\section{Numerical results}
\begin{figure*}[!t]
    \centering
        \subfloat[Non-i.i.d and equal user distributions.]{
            \includegraphics[width=0.75\columnwidth]{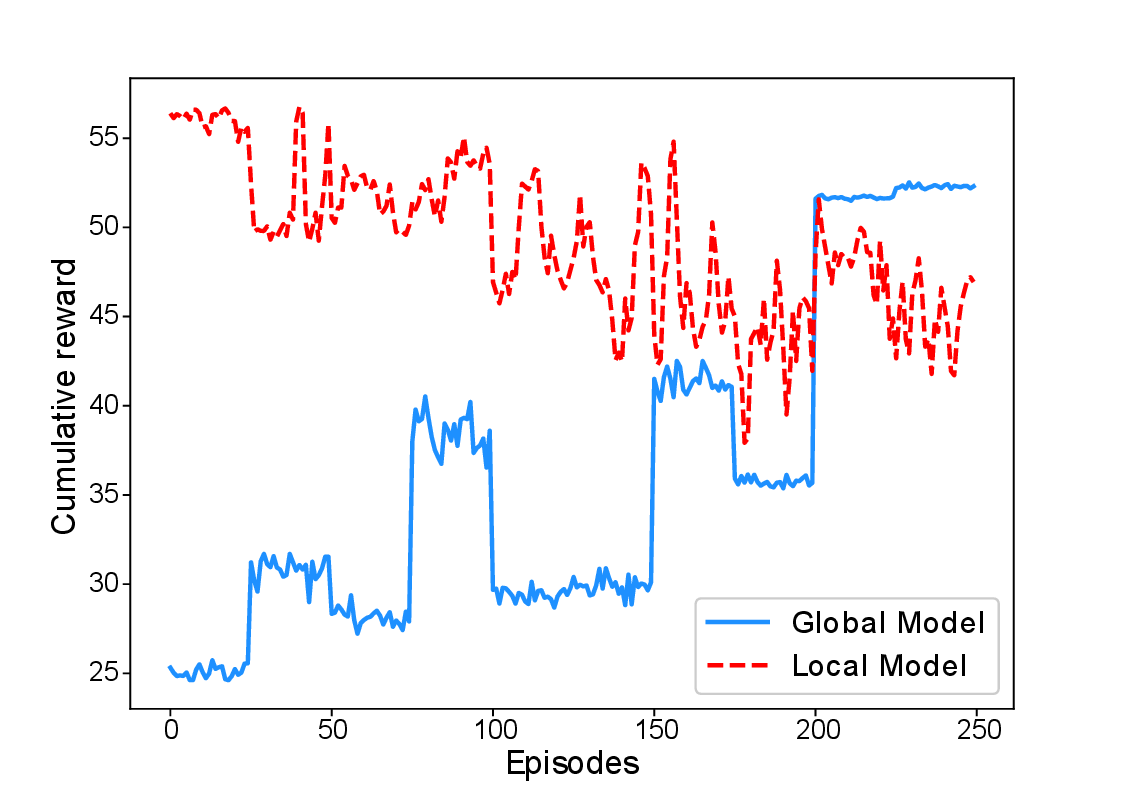}
            \label{fig:exp_noniid_equal}
        }
        \quad
        \subfloat[Non-i.i.d and unequal user distributions.]{
            \includegraphics[width=0.75\columnwidth]{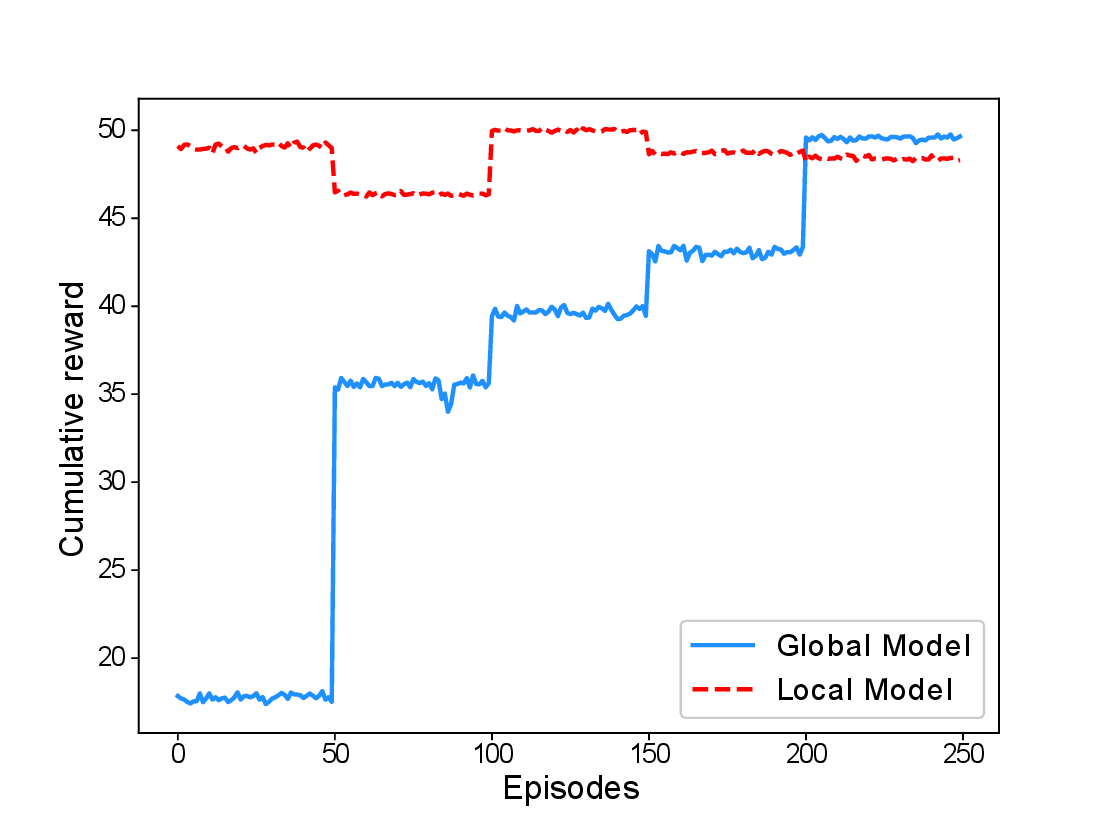}
            \label{fig:exp_noniid_unequal}
        }
    \caption{Training performance.}
    \label{fig:T_Comm}
\end{figure*}
\subsection{Experiment parameters and scenarios}
We consider a RIC-enabled RAN architecture with a single base station. The simulated users are randomly scattered in an area of $500m \times 500m$ around the BS, and served by 3 MVNOs.

The MVNOs collectively train a DDPG model.  The four networks of the model have two hidden fully connected layers with 400 and 300 neurons, respectively. We consider that each MVNO has a maximum number of users equal to 5. Since the maximum number of users is 5, the size of the input layer is 10 and the output layer is 5. 
We used the Rectified linear unit as an activation function since it avoids vanishing gradients in backpropagation, especially since the action space is limited to small values. This is due to one of the conditions stating that the allocated fraction should be less than $ 0.3$ to avoid squandering the bandwidth over a small handful of the users.
We used the Adam optimizer with two different learning rates for the actor and critic. 

We implemented the proposed architecture using the Python programming language. We used the PyTorch framework to implement the DDPG algorithm. The simulations were conducted on a laptop with a 2.6 GHz Intel i7 Processor, 16GB of RAM, and NVIDIA GeForce RTX 2070 graphic card.

For the FDRL mechanism, the training takes place over a total of 5 communication rounds. In each round, the model is trained by each MVNO for 500 episodes before sending the model to the RIC for aggregation. Each episode is composed of 50 steps, where the channel gains values are reset in each step, and the users' locations are reset every 25 episodes. In order to generate non-i.i.d distributions for user requirements, we set different probabilities of URLLC and eMBB users for each MVNO. The set of probabilities of URLLC users are $25\%$, $50\%$, and $75\%$ for MVNOs 1, 2, and 3, respectively. 
 To further test our proposed solution, we generated an unequal distribution of users. Specifically, we considered a case where MVNOs 1, 2, and 3 have 5, 4, and 3 users, respectively. In this case, the fraction of the bandwidth allocated to each MVNO is proportional to its number of users.

In the following, we study two scenarios: non-i.i.d with equal number of users, and non-i.i.d with unequal numbers of users. To evaluate the performance of FDRL, we compare the global model, \ie our model trained with FDRL, with the local models that were trained by each MVNO without collaboration. We first show the evolution of the training rewards, then we stress test the resulting models under different distribution shift scenarios to evaluate their robustness.
\subsection{FDRL training results}

The first considered scenario is non-i.i.d with equal number of users. The total number of users is 15, with 5 users being served by each MVNO. Fig. \ref{fig:exp_noniid_equal} shows the evolution of the average reward of the local models and the global model across 5 experiments. While the global model improves through the shared experience, even surpassing the local models average in later rounds, the local models have degrading performances throughout the training likely due to overfitting. In fact, since in later rounds, the exploration induced by the OU noise is reduced, the local models allocate less bandwidth to the users, which degrades the values of the rewards. In contrast, the global model learns slower to generalize, but achieves more robust training overall by leveraging the shared experience.  

\begin{figure*}[!t]
    \centering
  \begin{subfigure}{.35\linewidth}
    \centering
    \includegraphics[width = \linewidth]{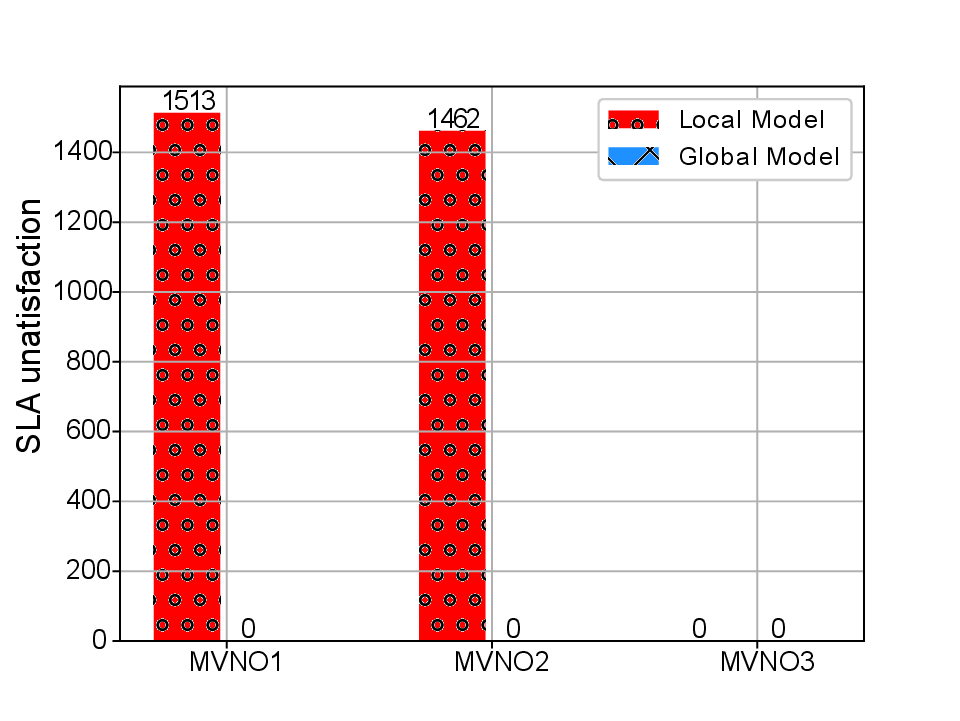}
    \caption{URLLC (75\%,25\%,50\%)}
    \label{fig:unvalid_delay_1_21}
  \end{subfigure}%
  \hspace{1em}
  \begin{subfigure}{.35\linewidth}
    \centering
    \includegraphics[width = \linewidth]{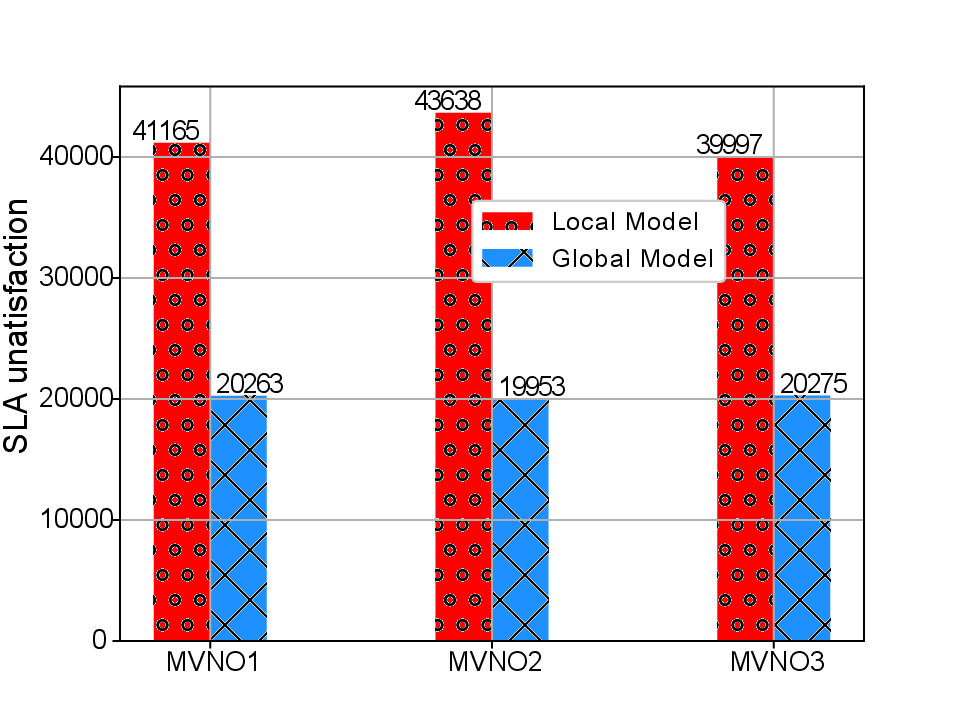}
    \caption{eMBB (75\%,25\%,50\%)}
    \label{fig:unvalid_rates_1_2}
  \end{subfigure}%
  
  \begin{subfigure}{.35\linewidth}
    \centering
    \includegraphics[width = \linewidth]{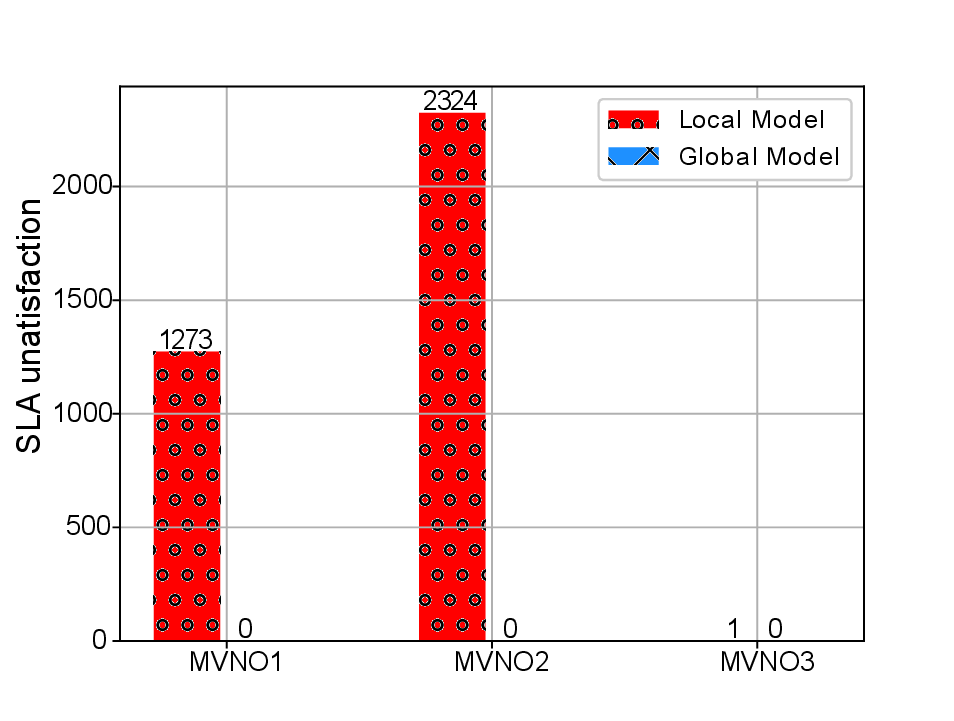}
    \caption{URLLC (50\%, 75\%,25\%)}
    \label{fig:unvalid_delay_1_21}
  \end{subfigure}%
  \hspace{1em}
  \begin{subfigure}{.35\linewidth}
    \centering
    \includegraphics[width = \linewidth]{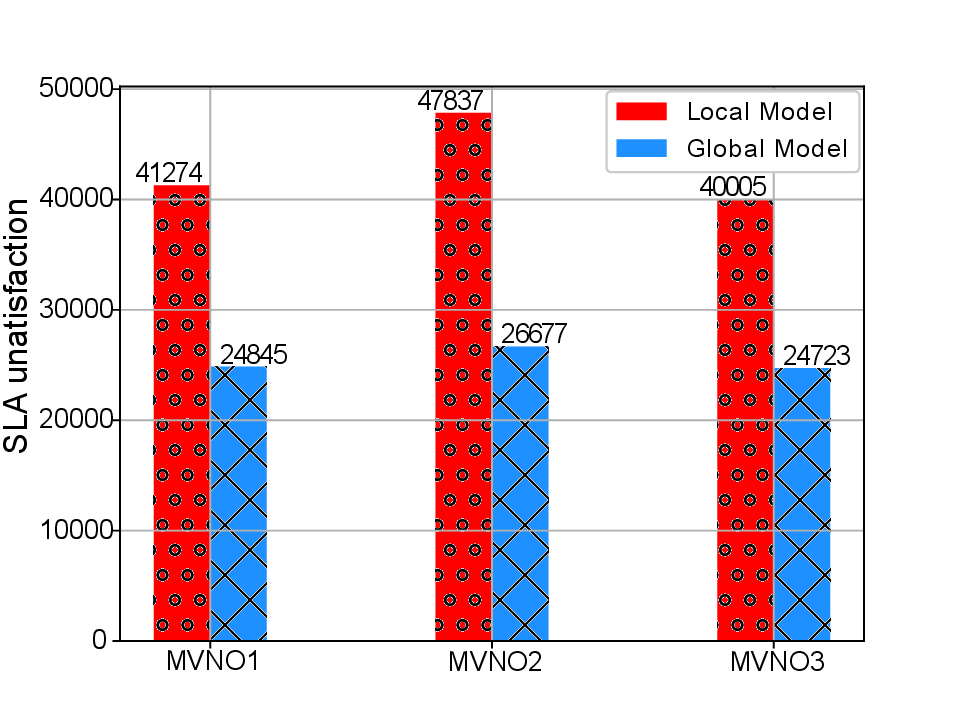}
    \caption{eMBB (50\%, 75\%,25\%)}
    \label{fig:unvalid_rates_1_2}
  \end{subfigure}%
  
  \caption{\label{fig:unvalid_exp_1}Evaluation under varying user distributions.}
\end{figure*}

The second considered scenario is non-i.i.d with unequal number of users. The total number of users is 12, where 5, 4, and 3 users are served by the first, second, and third MVNO, respectively. Fig. \ref{fig:exp_noniid_unequal} shows the evolution of the average reward of the local models and the global model across 5 experiments. Our first observation is that the cumulative rewards for both models are less than what was achieved in the case of equal numbers of users. This is mainly due to the punishment related to allocating bandwidth to non-existent users. Furthermore, similarly to the previous experiments, the global model improves slowly throughout the communication rounds, while the local models do not improve. 

\vspace{-0.4cm}
\subsection{FDRL performance evaluation}
To evaluate the performance of the proposed FDRL mechanism, we compared the number of invalid actions of the global model against each local model. Note that an action is considered invalid if it does not meet the user's SLA requirements. We use the resulting local models and global models, and test them in different environments by varying the underlying user types' distributions of each MVNO, then varying the number of users served by each MVNO.

\subsubsection{Varying user types' distributions}

The first considered scenario is non-i.i.d with equal number of users. The models are trained with a total of 15 users, where 5 users are served by each MVNO. To evaluate the robustness of the models in the case of changing users' requirements, we varied the underlying distributions of the users for each MVNO. 
The probabilities of URLLC users in the trained models are $25\%$, $50\%$, and $75\%$, for the first, second, and third MVNO, respectively.

In a first experiment, we changed the URLLC probabilities in test phase to $75\%$, $25\%$ ,$50\%$ for the first, second and third MVNOs, respectively.  In a second experiment, we changed these probabilities to $50\%$, $75\%$, $25\%$. 
Fig. \ref{fig:unvalid_exp_1}  shows the cumulative number of times where users' SLA requirements were not satisfied by the local models of the MVNOs and by the global model, while observing the same environments for a total of 20000 observations. We noticed that, overall, the global model's actions are less prone to violate the SLA requirements for eMBB and URLLC users compared to the individually trained models. Additionally, as we attributed larger weights to the URLLC users, the global model prioritizes this type of users and is less likely to violate their required delay. 

\subsubsection{Varying number of users}

\begin{figure*}[!t]
  \centering
  \begin{subfigure}{.35\linewidth}
    \centering
    \includegraphics[width = \linewidth]{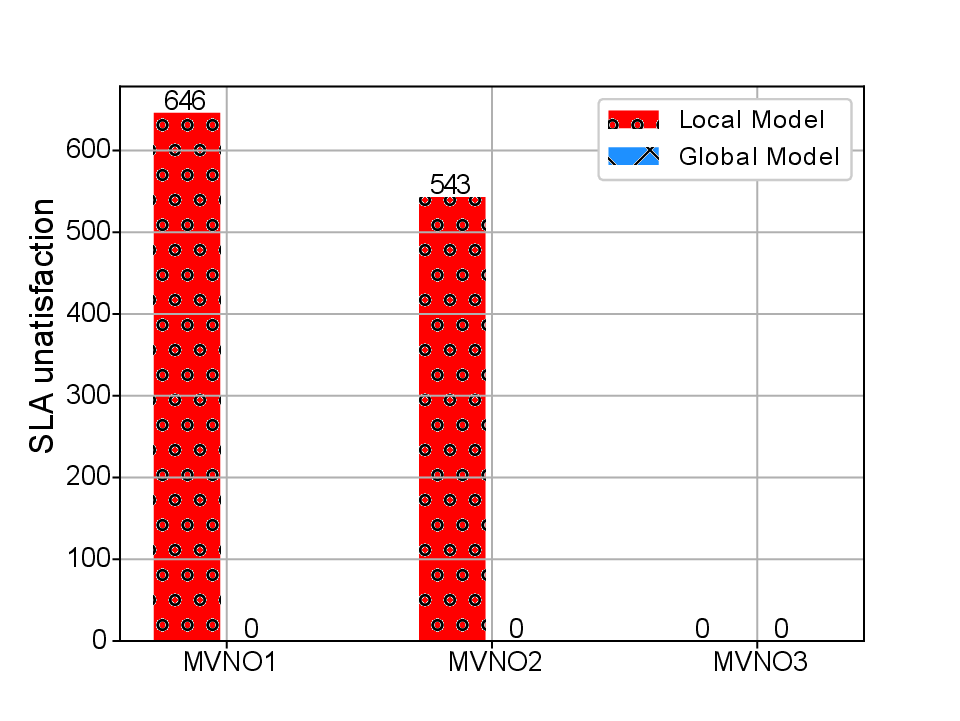}
    \caption{URLLC (4,3,5)}
    \label{fig:unvalid_delay_2_1}
  \end{subfigure}%
  \hspace{1em}
  \begin{subfigure}{.35\linewidth}
    \centering
    \includegraphics[width = \linewidth]{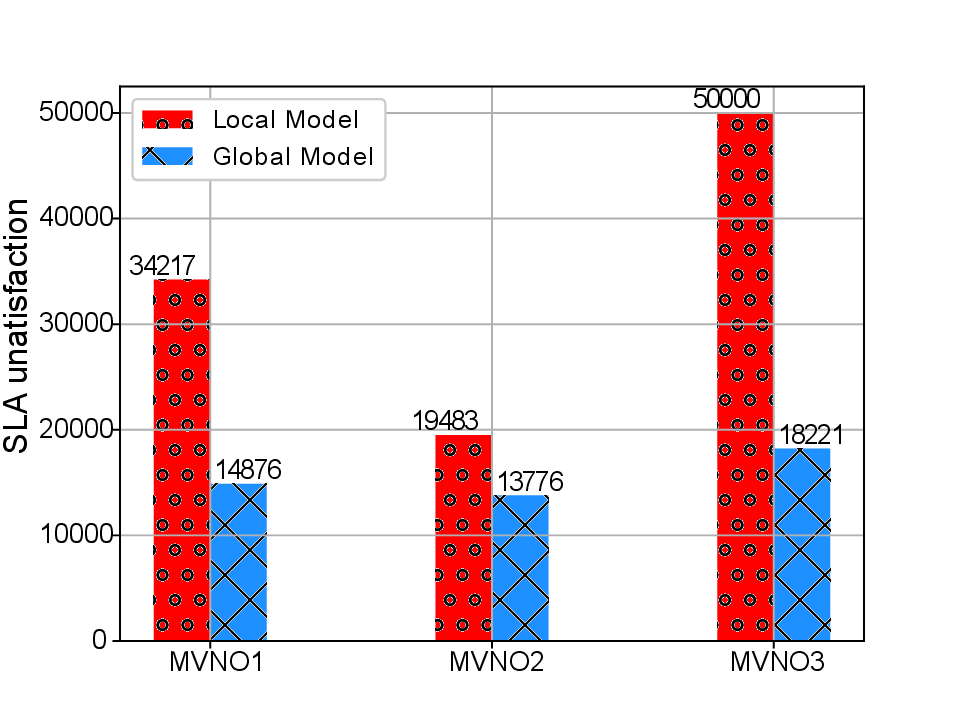}
    \caption{eMBB (4,3,5)}
    \label{fig:unvalid_rates_2_1}
  \end{subfigure}
  \hspace{1em}
  \begin{subfigure}{.35\linewidth}
    \centering
    \includegraphics[width = \linewidth]{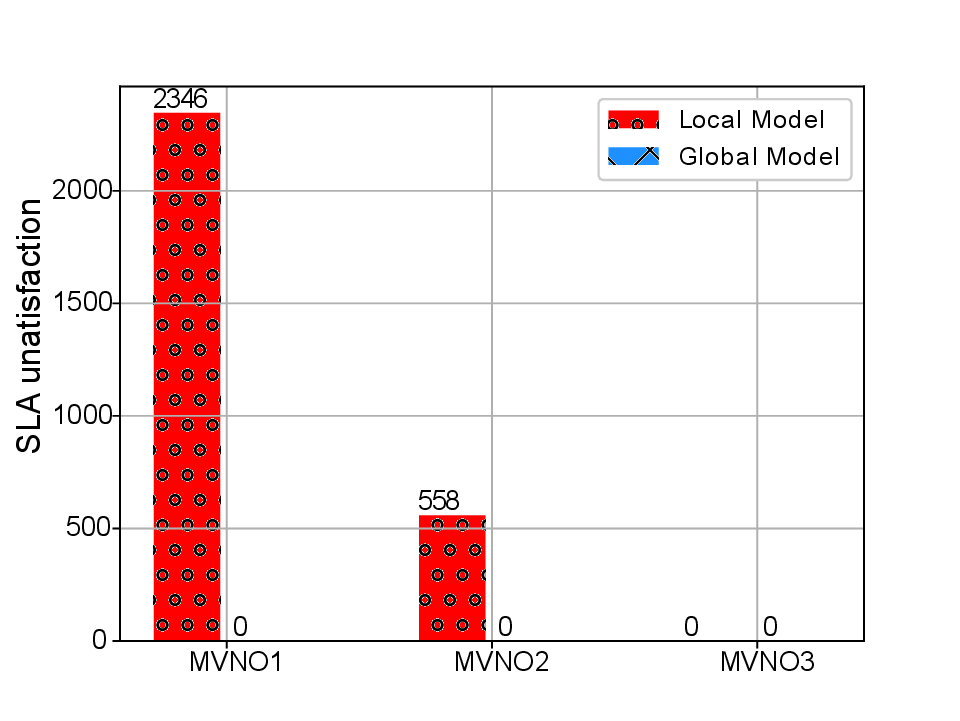}
    \caption{URLLC (3,5,4)}
    \label{fig:unvalid_delay_2_2}
  \end{subfigure}%
  \hspace{1em}
  \begin{subfigure}{.35\linewidth}
    \centering
    \includegraphics[width = \linewidth]{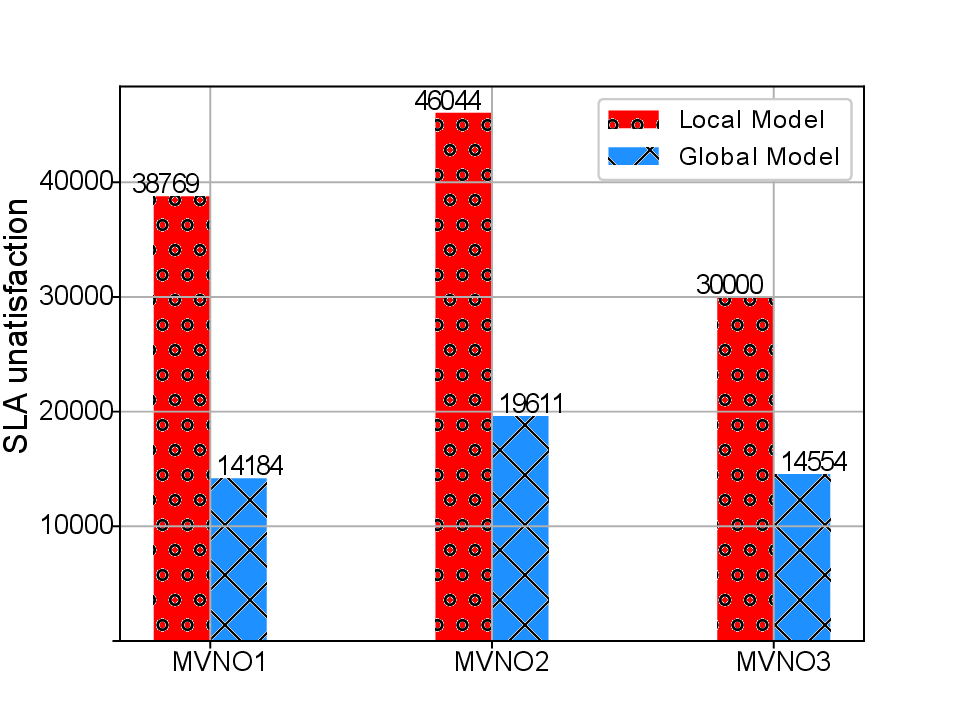}
    \caption{eMBB (3,5,4)}
    \label{fig:unvalid_rates_2_2}
  \end{subfigure}%
  
  \caption{\label{fig:unvalid_exp_2} Evaluation under different number of users.}
\end{figure*}

The second considered scenario is non-i.i.d with unequal number of users. The models are first trained with a total number of users of 12, where 5, 4, and 3 users are served by the first, second, and third MVNO, respectively. We seek to evaluate the robustness of the models in the case of changing number of users. 
First, we changed the number of users in test time to 4, 3 ,5 for the first, second and third MVNOs, respectively.  In a second experiment, we changed these numbers to 3, 5, 4. 
Fig. \ref{fig:unvalid_exp_2} shows the number of times where users' SLA were not satisfied by  the local models of the MVNOs and the global model, while observing the same environments for a total of 20000 observations. 

Similarly to the previous experiments, the global model's actions are less likely to violate the SLA requirements for eMBB and URLLC users compared to the individually trained models. Moreover, the third MVNO, being trained with mostly URLLC users, it has high satisfaction rate for this type, but it performs poorly for the eMBB users.  In general, the enhancement in the QoS for both types of users using the global model makes it worthwhile for the MVNOs to collaborate. 
\section{Conclusion and Future Work}
In this article, we proposed a FDRL RAN slicing mechanism in a multi-MVNOs environment to allocate radio resources to URLLC and eMBB users. We leverage FL to design a RAN slicing mechanism in which MVNOs collaborate to improve their policies for allocating radio resources to their users. We formulated the resource allocation problem for users of each MVNO as a single-agent MDP and used the DDPG algorithm to solve it. Then, FL is used to improve MVNOs' radio resource allocation models. The proposed FDRL mechanism has proven to be robust in meeting the diverse requirements of URLLC and eMMB users in different scenarios and more resilient to the environment changes.

Although the proposed FDRL-based RAN slicing mechanism allows a better robustness of radio resource allocation models, different interesting research directions deserve to be investigated. It is worth studying how to detect and track any deterioration in the performance of DRL models to proactively trigger the retraining process.  Another interesting issue to investigate is how to further enhance privacy beyond what federated learning inherently promotes. Indeed, while FL guarantees data privacy to a certain extent, it is still possible for adversaries to infer data from the model updates. To mitigate this, integrating other privacy preserving techniques such as differential privacy could be further explored in the case  of MNVOs cooperating in FDRL schemes.

\bibliographystyle{ieeetr}
\bibliography{IEEEabrv,refs}
\renewenvironment{IEEEbiography}[1]
  {\IEEEbiographynophoto{#1}}
  {\endIEEEbiographynophoto}

\vspace{-1.4cm}

\begin{IEEEbiography}{Amine Abouaomar} 
[M] is currently an assistant professor of computer science at Al Akhawayn University, Morocco. He was a Post-Doctoral Research Fellow at Polytechnique Montréal. He received his Ph.D. in electrical engineering from Université of Sherbrooke, Canada, and his Ph.D. in Computer Science from ENSIAS, Mohammed V University of Rabat, Morocco in 2021. He is an active IEEE member since 2017 and he is a technical chair program member of many international conferences. His research interest includes beyond 5G network, federated learning, and green machine learning.
\end{IEEEbiography}
\vspace{-1.4cm}
\begin{IEEEbiography}{Afaf Taïk}
[M] received her Ph.D. degree in Electrical Engineering from Université de Sherbrooke, Canada in 2022.  She holds a B.Eng. degree in software engineering from ENSIAS, Rabat, Morocco (2018), and a DESS in applied sciences from the Université de Sherbrooke, Canada (2018). She is the recipient of the best paper award from IEEE LCN 2021. Her research interests include edge computing and machine learning.
\end{IEEEbiography}
\vspace{-1.4cm}
\begin{IEEEbiography}{Abderrahime Filali}
[M] is currently working as a Post-Doctoral Research Fellow at the Department of Computer and Software Engineering at Polytechnique Montreal. He received his Ph.D. degree in electrical engineering from the Université de Sherbrooke, Canada in 2021. He served as a reviewer for several international conferences and journals. His current research interests include resource management in next-generation networks.
\end{IEEEbiography}
\vspace{-1.4cm}
\begin{IEEEbiography}{Soumaya Cherkaoui} [SM] is a Full Professor at the Department of Computer and Software Engineering at Polytechnique Montreal. She has been on the editorial board of several IEEE journals. She authored numerous conference and journal papers.  Her work was awarded with recognitions including several best paper awards at IEEE conferences. She is an IEEE ComSoc Distinguished Lecturer, a Professional Engineer in Canada, and has served as Chair of the IEEE ComSoc IoT-Ad Hoc and Sensor Networks Technical Committee.
\end{IEEEbiography}
\end{document}